\documentclass[11pt,a4paper]{article}
\pdfoutput=1
\usepackage{jheppub}
\usepackage{amsmath,amsfonts,amsthm,graphicx,ulem}
\usepackage{float}

\newcommand{\comment}[1]{}
\newcommand{\bal}{\begin{align}}
\newcommand{\eal}{\end{align}}
\newcommand{\beq}{\begin{equation}}
\newcommand{\eeq}{\end{equation}}
\newcommand\beqa{\begin{eqnarray}}
\newcommand\eeqa{\end{eqnarray}}
\newcommand\bea{\begin{array}}
\newcommand\eea{\end{array}}

\renewcommand{\geq}{\geqslant}

    \newcommand{\COMMENT}[1]{}
    
    \newcommand{\neqa}{\nonumber\end{eqnarray}}

\def\[{\left[}
\def\]{\right]}

\def\<{\langle}
\def\>{\rangle}

\def\i2{\frac{i}{2}}

\title{ \center{A new representation for two- and three-point correlators of operators from \(sl(2)\) sector}}
\author[a]{Evgeny Sobko}

\affiliation[a]{Ecole Normale Superieure, LPT,  75231 Paris CEDEX-5,
  France}

\emailAdd{evgenysobko AT gmail.com}

\abstract{We construct a new representation for two- and three-point correlators of operators from \(sl(2)\) sector of planar \(N=4\) SYM. The spin and twist of operators are arbitrary. We start with the correlation function of light-ray operators and carry out a projection to particular local operators using the method of Separated Variables. With the same calculation we obtain polynomials which are dual to wave functions of \(sl(2,R)\) spin-chain.}

\subheader{LPT ENS-13/22}

\usepackage{varioref}
\usepackage{makeidx}
\makeindex
\usepackage{amssymb}

\begin{document}

  \maketitle
\section{Introduction}
The operator product expansion in \(N\) = 4 SYM theory, as in any CFT , is completely
characterized by its 2-point and 3-point correlators, or, in other words, by the spectrum \(\Delta_j(\lambda)\) of
anomalous dimensions of local operators \(\mathcal{O}_j\) and by the structure constants\footnote{Which are tensors in the general case of operators with spin.} \(C_{ijk}(\lambda)\). Both the dimensions and the structure constants are in general complicated functions of coupling \(\lambda=g^2N_c\) and of quantum numbers of the operators. For the \(N\)=4 SYM
spectral problem, there has been a lot of progress in the last years \cite{Gromov:2009tv} allowing to study it
numerically, at any coupling. Recently, these developments have culminated in the formulation
of a well defined system of Riemann-Hilbert equations \cite{Gromov:2013pga}. However, for the correlation functions the
situation is far more complicated, and one is here in the early stage of a case-by-case study in
a weak or strong coupling regime.

A significant progress was achieved for \(su(2)\) sector in the weak-coupling regime. The method of "tailoring" of Bethe states which was proposed in \cite{Escobedo:2010xs} has been greatly evolved \cite{Foda:2011rr,Gromov:2011jh,Gromov:2012vu,Serban:2012dr,Kostov:2012jr,Kostov:2012yq,Gromov:2012uv,Vieira:2013wya}. It was applied to \(su(3)\) sector \cite{Foda:2013nua} as well as to higher loops in \(su(2)\) case \cite{Gromov:2012vu,Gromov:2012uv}. On the other hand, the case of noncomtact \(sl(2)\) sector has been much less investigated. Some interesting results concerning \(sl(2)\) sector one can find in \cite{Alday:2013cwa,Kazakov:2012ar,Georgiou:2011qk,Eden:2011we,Eden:2012tu,Eden:2012rr}. One can find in \cite{Alday:2013cwa} an interesting all-loop prediction for the case of two protected operators and one twist-2 operator with large spin. One loop prediction for two BPS and one \(sl(2)\) operator of arbitrary spin and twist is presented in \cite{Vieira:2013wya}.

In this note we propose a new approach to the calculation of correlation functions of \(sl(2)\) operators \(\mathcal{O}_{s,L}=\text{tr}D_+^sZ^L\) with arbitrary spin \(s\) and twist \(L\) in the leading order in the coupling. We are starting with the calculation of correlation function of nonlocal light-ray operators which serve as generating functions for local operators \(\mathcal{O}_{s,L}\). In order to make projection on particular local operators we use Sklyanin's method of Separated Variables (SoV)  \cite{Sklyanin:1991ss,Sklyanin:1995bm,Derkachov:2001yn,Derkachov:2002tf,Smirnov,Kazama:2013rya}. To generate wave-functions in SoV representation, we act a few times on the correlator of light-ray operators by \(Q\)-operator which was constructed in \cite{Derkachov:1999pz}. Further, we are using the scalar product \cite{Derkachov:2002tf} on the space of wave-functions in SoV representation in order to make projection on particular states. As a check for our method, we compare our formula in the case of twist-2 operators with a direct calculation involving explicit expression for the wave-function through Gegenbauer polynomials.

\section{Light-ray operator as a generating function}
In this section we collect some facts about light-ray operators. For more details, see review \cite{Belitsky:2004cz}.

\begin{figure}[H]
\center{\includegraphics[scale=1.4]{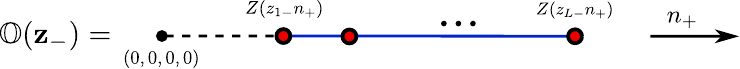} \\}
\caption{Graphical representation of the light-ray operator \(\mathbb{O}(\textbf{z}_-)\)}
\label{ris:LightRay}
\end{figure}

Let us introduce the light-ray operator \(\mathbb{O}(\textbf{z}_-)\):
\begin{gather}
\mathbb{O}(\textbf{z}_-)=\text{tr}\left(Z(n_+z_{1-})[n_+ z_{1-},n_+ z_{2-}]Z(n_+ z_{2-})...Z(n_+ z_{L-})[n_+ z_{L-},n_+ z_{1-}]\right), \label{DefNonLoc}
\end{gather}
where \(n_+\) is a light-ray vector\footnote{We use the basis \(\{n_+,n_-,e_{1\bot},e_{2\bot}\}\), where \(n_+^2=n_-^2=0\), \((n_+n_-=1)\) and any vector can be decomposed in the following way \(x=x_-n_++x_+n_-+x_\bot\).} \(n_+^2=0\), \(Z\) is a complex scalar field, \(\textbf{z}_-=\{z_{1-},...,z_{L-}\}\) is a set of coordinates of \(Z\)-fields along \(n_+\) direction and \([x,y]=\text{Pexp}\ ig_{{}_{YM}}\int\limits_{x}^y dz^\mu A_\mu(z)  \) is a gauge link between \(x\) and \(y\) (see in Fig.\ref{ris:LightRay}). We will further omit these gauge links, because we take into consideration only Born level approximation.

The operator \(\mathbb{O}(\textbf{z}_-)\) can be expanded in the Taylor series:
\begin{gather}
\mathbb{O}(\textbf{z}_-)=\text{tr}\left[\sum\limits_{i_1=0}^\infty z_{1-}^{i_1}\frac{D_+^{i_1}Z(0)}{i_1!}...\sum\limits_{i_L=0}^\infty z_{L-}^{i_L}\frac{D_+^{i_L}Z(0)}{i_L!}\right]=\notag\\
=\sum\limits_{i_1=0}^\infty...\sum\limits_{i_L=0}^\infty z_{1-}^{i_1}...z_{L-}^{i_L}\text{tr}\left[\frac{D_+^{i_1}Z(0)}{i_1!}...\frac{D_+^{i_L}Z(0)}{i_L!} \right],
\end{gather}
where \(D_+=n^\mu_+D_\mu\) is covariant derivative in \(n_+\) direction. This decomposition is an expansion over local operators. There is a distinguished basis of local operators. They diagonalize dilatation operator which has the form of \(sl(2,R)\) spin chain Hamiltonian in the one-loop approximation. All primary operators can be constructed using the Bethe ansatz technique. The full basis contains primaries as well as their descendants. Formally, the decomposition of  \(\mathbb{O}(\textbf{z}_-)\) reads as follows:
\begin{gather}
\mathbb{O}(\textbf{z}_-)=\sum\limits_{l,s,\{\alpha\}}\Phi_{l,s,\{\alpha\}}(\textbf{z}_-)\mathcal{O}_{l,s,\{\alpha\}}(0), \label{Decomposition over local}
\end{gather}
where \(\mathcal{O}_{l,s,\{\alpha\}}(0)\) is descendant of level \(l\) of primary operator which is characterised by the set of quantum numbers \((s,\{\alpha\})\). The natural label for primary operators is a set of Bethe roots or some functions of them. We explicitly labeled the  spin \(s\) hiding all other quantum numbers in \(\{\alpha\}\). The coefficient \(\Phi_{l,s,\{\alpha\}}(\textbf{z}_-)\) is a homogenous polynomial of order \(l+s\).
The local operators \(\mathcal{O}_{l,s,\{\alpha\}}(0)\) are in one to one correspondence with polynomials \(\Psi_{l,s,\{\alpha\}}(\textbf{p}_-)\):
\begin{gather}
\mathcal{O}_{l,s,\{\alpha\}}(0)=\bar{\Psi}_{l,s,\{\alpha\}}(\textbf{\(\partial_{\textbf{z}_-}\)})\mathbb{O}(\textbf{z}_-)|_{\textbf{z}_-=0}. \label{Psi}
\end{gather}
Explicit form of polynomials \(\Psi_{l,s,\{\alpha\}}(\textbf{p}_-)\) can be found by applying the Bethe ansatz technic.
All descendants have the following form:
\begin{gather}
\Psi_{l,s,\{\alpha\}}(\textbf{\(\partial_{\textbf{z}_-}\)})=(S^-)^l\Psi_{s,\{\alpha\}}(\textbf{\(\partial_{\textbf{z}_-}\)}),\\
\Phi_{l,s,\{\alpha\}}(\textbf{z}_-)=r_{l,s,\{\alpha\}}(S^+)^l\Phi_{s,\{\alpha\}}(\textbf{z}_-),
\end{gather}
where operators \(S^-=\sum\limits_{i=1}^LS^-_{i}\), \(S^+=\sum\limits_{i=1}^L S^+_{i}\) are defined as a sum of \(sl(2,R)\) generators in the spin-\(\frac{1}{2}\) representation:
\begin{gather}
S_n^+=z_{n-}^2\frac{\partial}{\partial z_{n-}}+z_{n-}, \ \ \ \ S_n^-=-\frac{\partial}{\partial z_{n-}}, \ \ \ \ S_n^0=z_{n-}\frac{\partial}{\partial z_{n-}}+\frac{1}{2}.
\end{gather}

The coefficient \(r_{l,s,\{\alpha\}}\) is a normalizing constant defined by the following condition:
\begin{gather}
\bar{\Psi}_{l,s,\{\alpha\}}(\textbf{\(\partial_{z_-}\)})\Phi_{l,s,\{\alpha\}}(\textbf{z}_-)|_{\textbf{z}=0}=\delta_{l,l'}\delta_{s,s'}\delta_{\{\alpha\},\{\alpha'\}}.
\label{Normalization}
\end{gather}
As it was established in the paper \cite{Derkachov:2013bda}, polynomials \(\Psi_{s,\{\alpha\}}(\textbf{p}_-)\) and \(\Phi_{s,\{\alpha\}}(\textbf{z}_-)\) are related by the dual symmetry\footnote{For descendants this duality reads as \(0=0\), because \(\Psi_{l>0,s,\{\alpha\}}(\textbf{p}_-)|_{p_i=z_i-z_{i+1}}\equiv 0\).}:
\begin{gather}
\Psi_{s,\{\alpha\}}(\textbf{p}_-)=\xi_{s,\{\alpha\}}\Phi_{s,\{\alpha\}}(\textbf{z}_-),\ \ \ p_{i-}=z_{i-}-z_{(i+1)-},\ \  z_{(i+L)-}=z_{i-},\\
\zeta_{s,\{\alpha\}}=\frac{i^s}{Q_{s,\{\alpha\}}(\frac{i}{2})}=\left(\prod\limits_{k=1}^s(\frac{1}{2}+iu_k)\right)^{-1},
\end{gather}
where \(\{u_k\}\) is the set of Bethe roots, and \(Q_{s,\{\alpha\}}(u)=\prod\limits_{k=1}^s(u-u_k)\) is Baxter's Q-function.
\subsection{Q-operator}

One can obtain the explicit form of the conformal operators \(O_{s,\{\alpha\}}\) and their anomalous dimensions \(\gamma_{s,\{\alpha\}}\) by diagonalizing the dilatation operator in the \(sl(2)\) sector of  \(N\)=4 SYM. Having used
 (\ref{Decomposition over local}) and (\ref{Psi}), this spectral problem can be reformulated as an eigenproblem for the Hamiltonian \(\mathbb{H}\) which acts \cite{Balitsky:1987bk} on the space of  polynomials \(\phi(\textbf{z}_-)\):
\begin{gather}
\mathbb{H}\Phi_{s,\{\alpha\}}(\textbf{z}_-)=\gamma_{s,\{\alpha\}}\Phi_{s,\{\alpha\}}(\textbf{z}_-),\label{HamNonloc}\\
\mathbb{H}=g^2(H_{12}+...+H_{L1}),\\
H_{i,i+1}\phi(z_{i-},z_{(i+1)-})=\notag\\
=\int\limits_0^1\frac{d\tau}{\tau}(2\phi(z_{i-},z_{(i+1)-})-\phi(z_{i-},(1-\tau)z_{(i+1)-}+z_{i-})-\phi((1-\tau)z_{i-}+\tau z_{(i+1)-},z_{(i+1)-})).\notag
\end{gather}
The Baxter approach to this eigenproblem is based on the existence of operator \(\tilde{\mathbb{Q}}(u)\) which depends on complex variable \(u\), acts on the space of polynomials, and satisfies a set of conditions:
\begin{itemize}
 \item \([\tilde{\mathbb{Q}}(u_1),\tilde{\mathbb{Q}}(u_2)]=0\),
 \item \([\tilde{\mathbb{Q}}(u_1),T(u_2)]=0\),
 \item \(\tilde{\mathbb{Q}}(u+i)(u+\frac{i}{2})^L+\tilde{\mathbb{Q}}(u-i)(u-\frac{i}{2})^L=T(u)\tilde{\mathbb{Q}}(u)\),
\end{itemize}
where \(T(u)=2u^L+q_2 u^{L-2}+...+q_L\) is the auxiliary transfer matrix for \(sl(2)\) spin chain, \(\{q_k\}\) is a complete set of commuting conserved charges which can serve as a label \((s,\{\alpha\})\) for states. Since such an operator exists, the problem (\ref{HamNonloc}) is equal to the diagonalization of \(\tilde{\mathbb{Q}}(u)\):
\begin{gather}
\tilde{\mathbb{Q}}(u)\Phi_{s,\{\alpha\}}(\textbf{z})=\frac{1}{c_{s,\{\alpha\}}}Q_{s,\{\alpha\}}(u)\Phi_{s,\{\alpha\}}(\textbf{z}_-),
\end{gather}
where \(Q_{s,\{\alpha\}}(u)\) is a Q-function defined above, and \(c_{s,\{\alpha\}}\) is a normalization constant which we fix as \(c_{s,\{\alpha\}}=\frac{i^s}{\zeta_{s,\{\alpha\}}}=Q_{s,\{\alpha\}}(\frac{i}{2})\).
The operator \(\tilde{\mathbb{Q}}(u)\), satisfying all conditions mentioned above, was constructed in \cite{Derkachov:1999pz}(see also \cite{Derkachov:2013bda}) and it reads as follows:
\begin{gather}
\tilde{\mathbb{Q}}(u)\phi(z_{1-},...,z_{L-})=[\Gamma(iu+\frac{1}{2})\Gamma(-iu+\frac{1}{2})]^{-L}\cdot\notag\\
\cdot\int\limits_0^1\prod\limits_{i=1}^L d\tau_i \tau_i^{-iu-1/2}(1-\tau_i)^{iu-1/2}\phi(\tau_1z_{1-}+(1-\tau_1)z_{2-},...,\tau_L z_{L-}+(1-\tau_L)z_{1-}).
\end{gather}
The operator \(\tilde{\mathbb{Q}}(u)\) is \(SL(2,R)\) invariant, and thus, we get the following action of \(\tilde{\mathbb{Q}}(u)\) on descendants:
\begin{gather}
\tilde{\mathbb{Q}}(u)\Phi_{l,s,\{\alpha\}}(\textbf{z}_-)=\frac{1}{c_{s,\{\alpha\}}}Q_{s,\{\alpha\}}(u)\Phi_{l,s,\{\alpha\}}(\textbf{z}_-).
\end{gather}
\section{Two-point correlation function}
Let us consider the correlator of two light-ray operators \(\mathbb{O}(x_0,\textbf{x}_-)\), \(\bar{\mathbb{O}}(y_0,\textbf{y}_-)\) stretched along \(n_+\) direction.  In the tree-level approximation both of them should have the same number \(L\) of fields \(Z\) and \(\bar{Z}\). Extra labels \(x_0\) and \(y_0\) indicate the starting points for these operators:
\begin{eqnarray}
x_0&=&(x_{0-},x_{0+},x_{0\bot}),\notag\\
y_0&=&(y_{0-},y_{0+},y_{0\bot}),\notag\\
x_i&=&(x_{0-}+x_{i-},x_{0+},x_{0\bot}),\notag\\
y_i&=&(y_{0-}+y_{i-},y_{0+},y_{0\bot}).
\end{eqnarray}
We fix the propagator for \(Z\)-field in the planar limit in the following way:
\begin{gather}
\langle Z^a_b(x) \bar{Z}^c_d(y) \rangle=\frac{\delta^a_d \delta^c_b}{N_c}\frac{1}{|x-y|^2}.
\end{gather}
The correlator in the tree-level approximation simply reads as follows:
\begin{gather}
\omega_L(x_0,\textbf{x}_-,y_0,\textbf{y}_-)=\langle \mathbb{O}(x_0,\textbf{x}_-)\bar{\mathbb{O}}(y_0,\textbf{y}_-) \rangle = \sum\limits_{\sigma} \prod\limits_{i=1}^{L}\frac{1}{|x_i-y_{\sigma(L+1-i)}|^2},\label{2pNonLoc}
\end{gather}
where \(\sigma\) is a cyclic permutation of \((1,...,L)\), and the sum goes over \(L\) different cyclic permutations.

On the other hand, one can expand \(\mathbb{O}(x_0,\textbf{x}_-)\), \(\bar{\mathbb{O}}(y_0,\textbf{y}_-)\) over local operators using  (\ref{Decomposition over local}), and rewrite the correlator of two nonlocal operators as a sum of 2-point correalators of local operators:
\begin{gather}
\langle\mathbb{O}(x_0,\textbf{x}_-)\bar{\mathbb{O}}(y_0,\textbf{y}_-)\rangle=\sum\limits_{l,s,\{\alpha\}}
\Phi_{l,s,\{\alpha\}}(\textbf{x}_-) \Phi_{l,s,\{\alpha\}}(\textbf{y}_-)\langle \mathcal{O}_{l,s,\{\alpha\}}(x_0) \bar{\mathcal{O}}_{l,s,\{\alpha\}}(y_0) \rangle. \label{2p cor decomp}
\end{gather}
Now let us act on \(\textbf{x}_-\) - coordinates by \(\tilde{\mathbb{Q}}_{\textbf{x}_-}(u)\)\footnote{Extra label "\(\textbf{x}_-\)" was introduced to stress that this operator acts on coordinates \(\textbf{x}_-=\{x_{1-},...,x_{L-}\}\). } on both sides of (\ref{2p cor decomp}):
\begin{gather}
\tilde{\mathbb{Q}}_{\textbf{x}_-}(u)\langle\mathbb{O}(x_0,\textbf{x}_-)\bar{\mathbb{O}}(y_0,\textbf{y}_-)\rangle=\notag\\
=\sum\limits_{l,s,\{\alpha\}}
\frac{Q_{s,\{\alpha\}}(u)}{c_{s,\{\alpha\}}}\Phi_{l,s,\{\alpha\}}(\textbf{x}_-) \Phi_{l,s,\{\alpha\}}(\textbf{y}_-)\langle \mathcal{O}_{l,s,\{\alpha\}}(x_0) \bar{\mathcal{O}}_{l,s,\{\alpha\}}(y_0) \rangle.
\end{gather}
Applying \(L-1\) operators \(\tilde{\mathbb{Q}}_{\textbf{x}_-}(u_1)...\tilde{\mathbb{Q}}_{\textbf{x}_-}(u_{L-1})\) to both sides of (\ref{2p cor decomp}), we get:
\begin{gather}
\prod\limits_{i=1}^{L-1}\tilde{\mathbb{Q}}_x(u_i)\langle\mathbb{O}(x_0,\textbf{x}_-)\bar{\mathbb{O}}(y_0,\textbf{y}_-)\rangle=\notag\\
=\sum\limits_{l,s,\{\alpha\}}
\prod\limits_{i=1}^{L-1}\frac{Q_{s,\{\alpha\}}(u_i)}{c_{s,\{\alpha\}}}\Phi_{l,s,\{\alpha\}}(\textbf{x}_-) \Phi_{l,s,\{\alpha\}}(\textbf{y}_-)\langle \mathcal{O}_{l,s,\{\alpha\}}(x_0) \bar{\mathcal{O}}_{l,s,\{\alpha\}}(y_0) \label{delaem Sklyanina}\rangle.
\end{gather}

Now let us introduce \(\Omega_{s,\{\alpha\}}(\textbf{u})=\Omega_{s,\{\alpha\}}(u_1,...,u_{L-1})\) - the wave function in the Sklyanin's Separated Variables(SoV):
\begin{gather}
\Omega_{s,\{\alpha\}}(\textbf{u})=\prod\limits_{k=1}^{L-1}Q_{s,\{\alpha\}}(u_k).
\end{gather}
The SoV representation for the \(sl(2,R)\) spin chain was constructed in \cite{Derkachov:2002tf}. The authors have explicitly established  unitary transformation to Separated Variables along with the Sklyanin's measure defining the scalar product in
the SoV representation. They have also proved equivalence of SoV and ABA methods.

The orthogonality condition for the wave functions in the SoV representation reads as follows:
\begin{gather}
\langle (s,\{\alpha\})_\textbf{u}|(s',\{\alpha'\})_\textbf{u} \rangle_{\tilde{\mu}}=\int\limits_{R^{L-1}}d^{L-1}\textbf{u}\,\tilde{\mu}(\textbf{u})\prod\limits_{k=1}^{L-1}Q_{s,\{\alpha\}}(u_k)Q_{s',\{\alpha'\}}(u_k)=N_{s,\{\alpha\}}\delta_{s,s'}\delta_{\{\alpha\},\{\alpha'\}},
\end{gather}
where \(|(s,\{\alpha\})_\textbf{u}\rangle=\Omega_{L,(s,\{\alpha\})}(\textbf{u})\), \(N_{s,\{\alpha\}}\) is a coefficient, and the label \(\tilde{\mu}\) means that the scalar product is defined by the measure \(\tilde{\mu}(\textbf{u})\), which has the following form:
\begin{gather}
\tilde{\mu}(\textbf{u})=\prod \limits_{\substack{j,k=1\\ j<k}}^{L-1}(u_k-u_j)\sinh(\pi(u_k-u_j))\prod\limits_{k=1}^{L-1}[\Gamma(\frac{1}{2}+iu_k)\Gamma(\frac{1}{2}-iu_k)]^L.\label{measure}
\end{gather}
Now let us obtain 2-point correlator of particular operators from the correlator of two light-ray operators (\ref{2p cor decomp}).

We are interested in the particular primary operator \(\mathcal{O}_{s,\{\alpha\}}\) with spin \(s\). Thus, we can expand \(\omega(x_0,\textbf{x}_-,y_0,\textbf{y}_-)\) in the series and collect the terms, such as \(P_s(\textbf{x}_-)Q_s(\textbf{y}_-)\) , where \(P_s(\textbf{x}_-)\) and \(Q_s(\textbf{y}_-)\) are homogenous polynomials of order \(s\). It can be easily done by one extra integration. Namely, one can replace \(\textbf{x}_-=(x_{1-},...,x_{L-})\) by rescaled coordinates \(\eta_\textbf{x}\textbf{x}_-=(\eta_\textbf{x} x_{1-},...,\eta_\textbf{x} x_{L-})\), and carry out contour integration \(\frac{1}{2\pi i}\oint\limits_{0}d \eta_\textbf{x}\frac{1}{\eta_\textbf{x}^{s+1}}(...)\) around zero. We introduce \(\omega_{L}^s(x_0,\textbf{x}_-,y_0,\textbf{y}_-)\), the projection of the function \(\omega_L(x_0,\textbf{x}_-,y_0,\textbf{y}_-)\) on the states with spin \(s\):
\begin{gather}
\omega_{L}^s(x_0,\textbf{x}_-,y_0,\textbf{y}_-)=\frac{1}{(2\pi i)^2}\oint\limits_{0}d \eta_\textbf{x}\frac{1}{\eta_\textbf{x}^{s+1}}\oint\limits_{0}d \eta_\textbf{y}\frac{1}{\eta_\textbf{y}^{s+1}}\omega(x_0,\eta_\textbf{x}\textbf{x}_-,y_0,\eta_\textbf{y}\textbf{y}_-).\label{2pSpinS}
\end{gather}
This projection corresponds to the contribution of all operators with spin \(s\). In the general case of arbitrary twist \(L\) we have several primary operators of the spin \(s\). Moreover, the descendants \(\Psi_{k,s-k,\{\alpha\}}\) also have spin \(s\) and contribute to (\ref{2pSpinS}). To separate one particular primary operator with quantum numbers \((s,\{\alpha\})\) we use orthogonality of the wave-functions in the SoV representation. As a first step, we generate wave functions \(\Omega_{s,\{\alpha\}}(\textbf{u})\), \(\Omega_{s,\{\alpha\}}(\textbf{v})\),  acting on \(\omega_{L}^s(x_0,\textbf{x}_-,y_0,\textbf{y}_-)\) by operators \(\tilde{\mathbb{Q}}_{\textbf{x}_-}(\textbf{u})=\prod\limits_{i=1}^{L-1}\tilde{\mathbb{Q}}_{\textbf{x}_-}(u_i)\), \(\tilde{\mathbb{Q}}_{\textbf{y}_-}(\textbf{v})=\prod\limits_{i=1}^{L-1}\tilde{\mathbb{Q}}_{\textbf{y}_-}(v_i)\):
\begin{gather}
\tilde{\mathbb{Q}}_{\textbf{x}_-}(\textbf{u})\tilde{\mathbb{Q}}_{\textbf{y}_-}(\textbf{v}) \omega_{L}^s(x_0,\textbf{x}_-,y_0,\textbf{y}_-)=\notag\\
=\sum\limits_{\substack{l,s',\{\alpha\}\\ l+s'=s}}\frac{1}{c_{s',\{\alpha\}}^{2L-2}}\Omega_{s',\{\alpha\}}(\textbf{u})\Omega_{s',\{\alpha\}}(\textbf{v})
\Phi_{l,s',\{\alpha\}}(\textbf{x}_-) \Phi_{l,s',\{\alpha\}}(\textbf{y}_-)\langle \mathcal{O}_{l,s',\{\alpha\}}(x_0) \bar{\mathcal{O}}_{l,s',\{\alpha\}}(y_0)\rangle,
\end{gather}
and then we use orthogonality:
\begin{gather}
\Phi_{s,\{\alpha\}}(\textbf{x}_-) \Phi_{s,\{\alpha\}}(\textbf{y}_-)\langle \mathcal{O}_{s,\{\alpha\}}(x_0) \bar{\mathcal{O}}_{s,\{\alpha\}}(y_0)\rangle=\notag\\
=\frac{c_{s,\{\alpha\}}^{2L-2}}{N_{s,\{\alpha\}}^2}\langle (s,\{\alpha\})_\textbf{u},(s,\{\alpha\})_\textbf{v}|\tilde{\mathbb{Q}}_{\textbf{x}_-}(\textbf{u})\tilde{\mathbb{Q}}_{\textbf{y}_-}(\textbf{v})\omega_{L}^s(x_0,\textbf{x}_-,y_0,\textbf{y}_-)\rangle_{\tilde{\mu}}, \label{2-point}
\end{gather}
where \(\langle (s,\{\alpha\})_\textbf{u},(s,\{\alpha\})_\textbf{v}|=\Omega_{s,\{\alpha\}}(\textbf{u})\Omega_{s,\{\alpha\}}(\textbf{v})\).

\subsection{Discussion of (\ref{2-point})}

\ \ \ \ \ \ At first, we should stress that the representation (\ref{2-point}) gives us in one calculation both 2-point correlator and the polynomial \(\Phi_{s,\{\alpha\}}\) which is dual to the wave function \(\Psi_{s,\{\alpha\}}\).

The second comment concerns normalization. One can multiply functions \(\Phi_{s,\{\alpha\}}\) by any constant \(c\) and, at the same time, multiply two-point correlator by \(\frac{1}{c^2}\). Thus, the left-hand side of (\ref{2-point}) will not be changed. This freedom in the normalization is not surprising, because we have fixed only the action of \(\Psi_{s,\{\alpha\}}(\partial_{\textbf{x}_-})\) on \(\Phi_{s,\{\alpha\}}(\textbf{x}_-)\) in the (\ref{Normalization}).

In order to obtain the two-point correlator of particular operators, it is sufficient to act in (\ref{2-point}) just by one operator \(\tilde{\mathbb{Q}}_{\textbf{x}_-}\), and take the scalar product with \(\langle(s,\{\alpha\})_\textbf{u}|=\Omega_{s,\{\alpha\}}(\textbf{u})\). Indeed, all other terms disappear due to orthogonality of local operators. Nevertheless, we choose the form as in (\ref{2-point}) because it is symmetric and well adopted for the normalization of three-point correlation functions.

Now let us notice, that the product of \(\Gamma\)-functions in the measure (\ref{measure}) is exactly canceled by \(\Gamma\)-functions which comes from operator \(\tilde{\mathbb{Q}}_{\textbf{x}_-}(\textbf{u})\). For this reason we introduce a new operator \(\mathbb{Q}_{\textbf{x}_-}(u)=[\Gamma(iu+\frac{1}{2})\Gamma(-iu+\frac{1}{2})]^{L}\tilde{\mathbb{Q}}_{\textbf{x}_-}(u)\). The action of operator \(\mathbb{Q}_{\textbf{x}_-}(\textbf{u})=\prod\limits_{i=1}^{L-1}\mathbb{Q}_{\textbf{x}_-}(u_i)\) implies \(L-1\) substitutions such as \(\{x_{j-}\rightarrow x_{j-}\tau_{ij}+x_{(j+1)-}(1-\tau_{ij})\}\). Nevertheless, this action can be explicitly formulated:
\begin{gather}
\mathbb{Q}_{\textbf{x}_-}(\textbf{u})\phi(\textbf{x}_-)=\prod\limits_{i=1}^{L-1}\mathbb{Q}(u_i)\phi(x_{1-},...,x_{L-})=
\int\limits_0^1\prod\limits_{i=1}^{L-1}\prod\limits_{j=1}^{L}d\tau_{ij} \tau_{ij}^{-iu_i-\frac{1}{2}}(1-\tau_{ij})^{iu_i-\frac{1}{2}}\phi(\hat{x}_{1-},...,\hat{x}_{L-}),\label{ActionQ}
\end{gather}
where \(\hat{x}_{l-}\) is a linear combination of all coordinates \(x_{k-}\). Operation "hat" in  \(\hat{x}_{l-}\) has an elegant graphical representation. 
\par\medskip
 \begin{figure}[H]
\center{\includegraphics[scale=0.7]{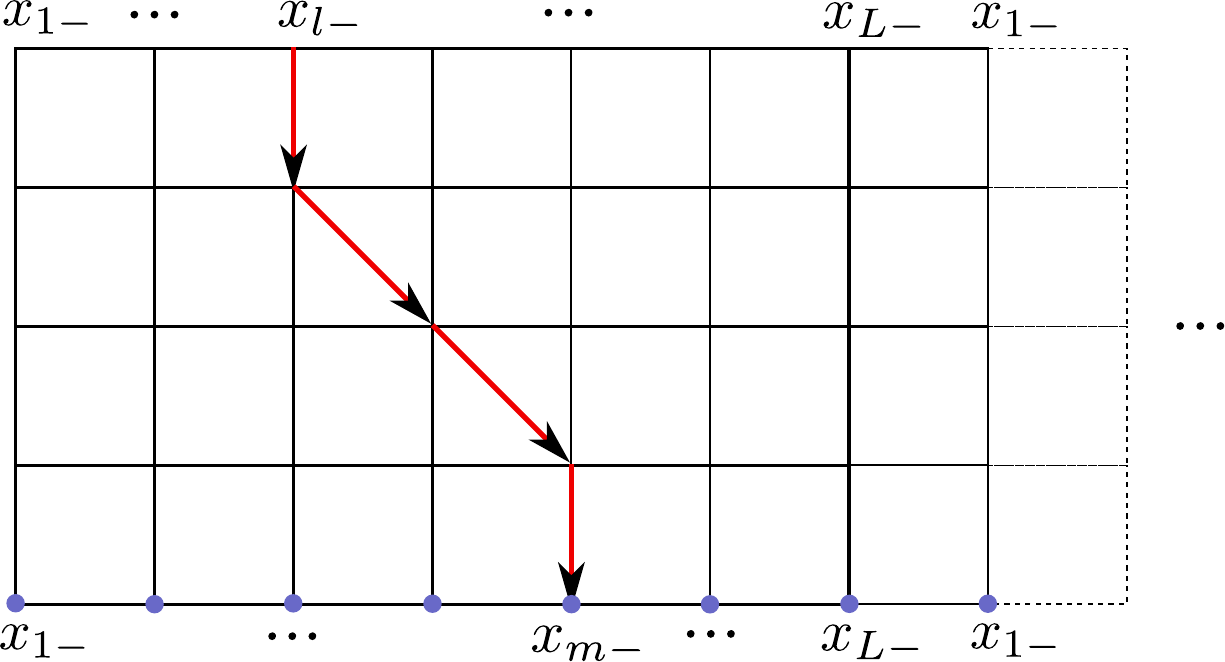} \\}
\caption{An example of path which contribute to \(\hat{x}_{l-}\).}
\label{ris:Path}
\end{figure}
\begin{figure}[H]
\center{\includegraphics[scale=0.5]{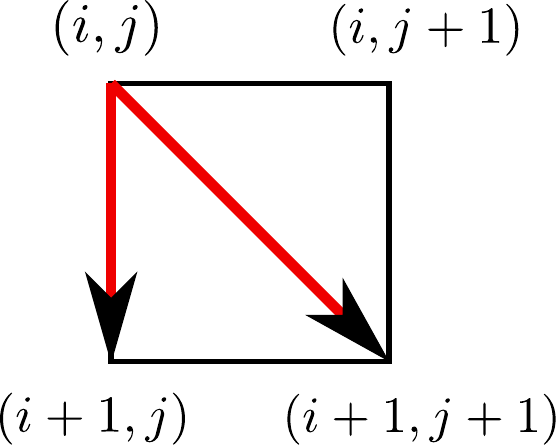} \\}
\caption{Path goes only from Nord to South or from Nord-West to South-East. First type of links gives us the multiplier \(\tau_{ij}\) while the second one gives us \((1-\tau_{ij})\).}
\label{ris:2Directions}
\end{figure}
\begin{figure}[H]
\center{\includegraphics[scale=0.6]{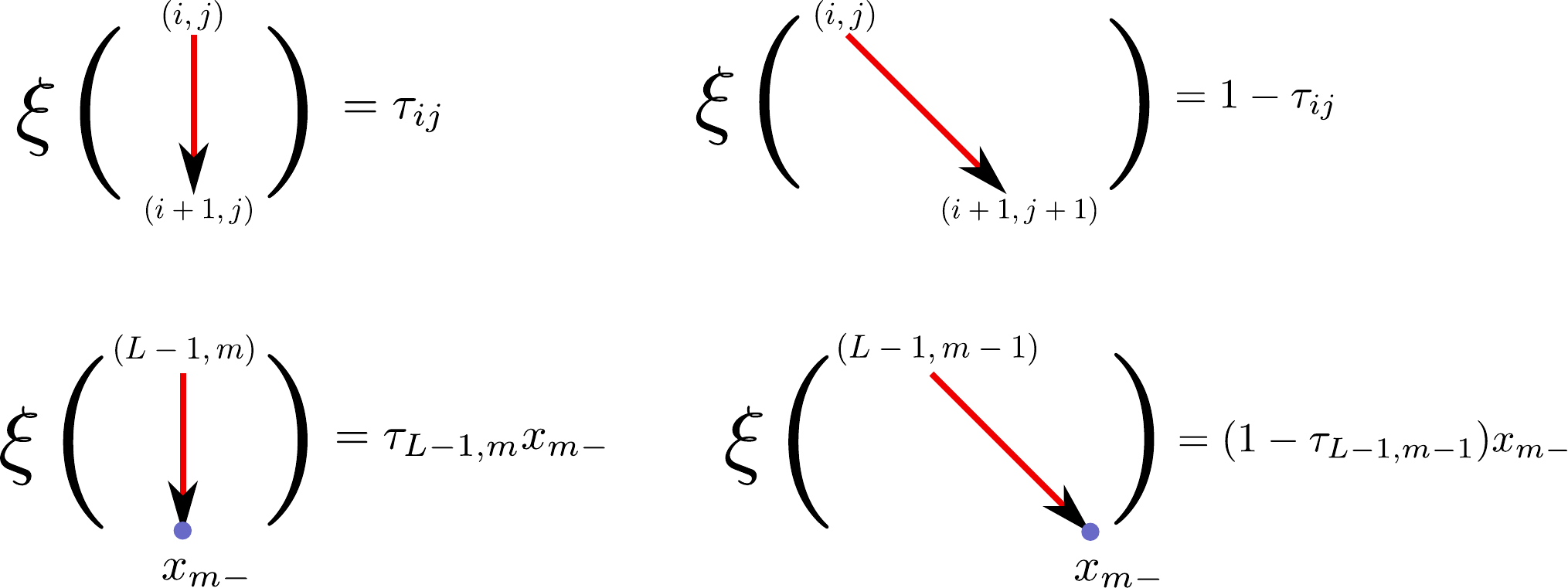} \\}
\caption{Action of function \(\xi\) on different arrows.}
\label{ris:Mapping}
\end{figure}

Let us define a strip of width \(L-1\) in the vertical direction and infinite in the horizontal direction with factorization \(x_{(l+L)-}=x_{l-}\). Formally, it is equivalent to the cylinder. All paths start on the upper horizontal boundary and go to the bottom line as depicted in the Fig.\ref{ris:Path}. Any path is a sequence of vertical and tilted arrows, as shown in the Fig.\ref{ris:2Directions}
Now we define function \(\xi\) which maps each path to the expression depending on \(\{\tau_{ij}\}\) and one of the coordinates \(\{x_{i-}\}\). Any path is a link of vertical and tilted arrows. The function \(\xi\) acts multiplicatively, namely, if the path is represented as a sequens of arrows \(abc...\) then \(\xi(abc...)=\xi(a)\xi(b)\xi(c)...\). The action of \(\xi\) on different arrows is represented in the Fig.\ref{ris:Mapping}
The expression for the \(\hat{x}_{l-}\) is given by the sum \(\hat{x}_{l-}=\sum\limits_{\zeta \in H_{x_{l-}}}\xi(\zeta)\) over the set \(H_{x_{l-}}\) of different paths of length \(L-1\) with the starting point \(x_{l-}\) on the upper boundary, and a final point on the down boundary.

The formula (\ref{2-point}) can be rewritten in the following way:
\begin{gather}
\Phi_{s,\{\alpha\}}(\textbf{x}_-) \Phi_{s,\{\alpha\}}(\textbf{y}_-)\langle \mathcal{O}_{s,\{\alpha\}}(x_0) \bar{\mathcal{O}}_{s,\{\alpha\}}(y_0)\rangle=\notag\\
=\frac{c_{s,\{\alpha\}}^{2L-2}}{N_{s,\{\alpha\}}^2}\langle (s,\{\alpha\})_{\textbf{u}},(s,\{\alpha\})_{\textbf{v}} |\mathbb{Q}_{\textbf{x}_-}(\textbf{u})\mathbb{Q}_{\textbf{y}_-}(\textbf{v})\omega_{L}^s(x_0,\textbf{x}_-,y_0,\textbf{y}_-)\rangle_\mu , \label{2-pointFinal}
\end{gather}
where operators \(\mathbb{Q}_{\textbf{x}_-}(\textbf{u})\) and \(\mathbb{Q}_{\textbf{y}_-}(\textbf{v})\) act as in the (\ref{ActionQ}), and \(\langle ...|... \rangle_\mu\) is defined as follows:

\begin{gather}
\langle f_1(\textbf{u})|f_2(\textbf{u})\rangle_\mu=\int\limits_{R^{L-1}}d^{L-1}\textbf{u}\mu(\textbf{u})f_1(\textbf{u})f_2(\textbf{u}),
\end{gather}
\begin{gather}
\mu(\textbf{u})=\prod \limits_{j,k=1; j<k}^{L-1}(u_k-u_j)\sinh(\pi(u_k-u_j)). \label{MeasurFin}
\end{gather}

\section{Three-point correlator}
\begin{figure}[H]
\center{\includegraphics[scale=1.1]{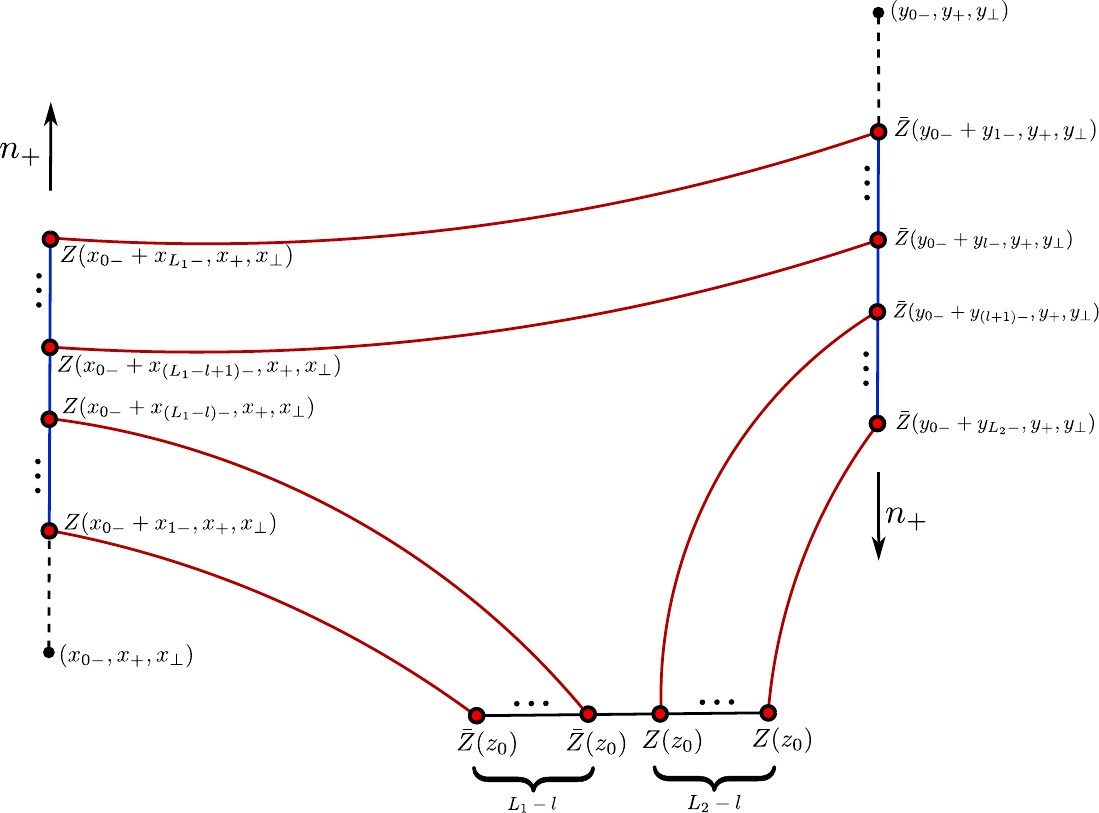} \\}
\caption{Three-point correlator of two nonlocal and one local operator in the tree-level approximation.}
\label{ris:3point}
\end{figure}

In this section we provide expressions for the three-point correlators , where a few of operators belong to \(sl(2)\) sector. Let's start with the case of two \(sl(2)\) operators \(\mathcal{O}_{L_1,(s_1,\{\alpha_1\})}=\text{tr}D_+^{s_1}Z^{L_1}\), \(\mathcal{O}_{L_2,(s_2,\{\alpha_2\})}=\text{tr}D_+^{s_2}\bar{Z}^{L_2}\), and one operator \(\mathcal{O}_{\triangle}=\text{tr}Z...\bar{Z}...\) with \(L_2-l\) fields \(Z\) and \(L_1-l\) fields \(\bar{Z}\). Our approach can be applied in a similar way to \(sl(2)\) operators with different polarizations, but for sake of brevity in notations we restrict ourselves to the case of one polarization \(n_+\).  The operator \(\mathcal{O}_\triangle\) consists of terms with different order of fields, and is mixed with fermions and gluons. In the leading order in the coupling\footnote{The planar limit is implied.}, nonzero contribution comes only from one term, when all fields \(Z\) are grouped on the left hand side, and all fields \(\bar{Z}\) on the right. Let's start with the correlation function of two nonlocal operators \(\mathbb{O}_Z(x_0,\textbf{x}_-)\), \(\mathbb{O}_{\bar{Z}}(y_0,\textbf{y}_-)\) stretched along \(n_+\)-direction and local operator \(\mathcal{O}_\triangle(z_0)\). The operator  \(\mathbb{O}_Z(x_0,\textbf{x}_-)\) consists of \(L_1\) fields \(Z\), the operator \(\mathbb{O}_{\bar{Z}}(y_0,\textbf{y}_-)\) consists of \(L_2\) fields \(\bar{Z}\). The correlator in the planar limit is depicted in Fig.\ref{ris:3point}
As we noticed before, only one term from the operator \(\mathcal{O}_\triangle\) gives nonzero contrubution to the correlator in the lowest order in the coupling. This term has the form \(c_{\triangle}\text{tr}Z^{L_2-l}\bar{Z}^{L_1-l}\), where \(c_{\triangle}\) is a coefficient. In this paper we are concentrated on the \(sl(2)\) operators. Due to this reason, we introduce extra normalisation \(\frac{1}{\mathcal{N}}\) which cancels this coefficient, irrelevant to our discussion. We will specify \(\mathcal{N}\) a bit later.

  The 3-point correlator can be calculated in the same way as in the 2-point case. Its expression reads as follows\footnote{To distinguish two operators of length \(L_1\) and \(L_2\), we explicitly introduce corresponding label. For example, polynomial \(\Phi_{L_1,(s_1,\{\alpha_1\})}(\textbf{x}_-)\) corresponds to the polynomial \(\Phi_{(s_1,\{\alpha_1\})}(\textbf{x}_-)\) of operator with twist \(L_1\), etc.}:
\begin{gather}
\Phi_{L_1,(s_1,\{\alpha_1\})}(\textbf{x}_-) \Phi_{L_2,(s_2,\{\alpha_2\})}(\textbf{y}_-)\langle \mathcal{O}_{L_1,(s_1,\{\alpha_1\})}(x_0) \mathcal{O}_{L_2,(s_2,\{\alpha_2\})}(y_0) \mathcal{O}_\triangle(z_0)\rangle=\notag\\
=H\langle \left(L_1,(s_1,\{\alpha_1\})\right)_{\textbf{u}},\left(L_2,(s_2,\{\alpha_2\})\right)_{\textbf{v}} |\mathbb{Q}_{\textbf{x}_-}(\textbf{u})\mathbb{Q}_{\textbf{y}_-}(\textbf{v})\omega^{s_1,s_2,0}_{L_1,L_2,M}(x_0,\textbf{x}_-,y_0,\textbf{y}_-;z_0)\rangle_\mu, \label{3-point}
\end{gather}
where \(H=\frac{1}{N_c}c_{\triangle}\frac{c_{L_1,(s_1,\{\alpha_1\})}^{L_1-1}c_{L_2,(s_2,\{\alpha_2\})}^{L_2-1}}{N_{L_1,(s_1,\{\alpha_1\})}N_{L_2,(s_2,\{\alpha_2\})}}\),
\begin{gather}
\langle(L_1,(s_1,\{\alpha_1\}))_{\textbf{u}},(L_2,(s_2,\{\alpha_2\}))_{\textbf{v}}|=\Omega_{L_1,(s_1,\{\alpha_1\})}(\textbf{u})\Omega_{L_2,(s_2,\{\alpha_2\})}(\textbf{v}),\notag\\
\langle...\rangle_\mu=\int\limits_{R^{L_1-1}}\int\limits_{R^{L_2-1}}d^{L_1-1}\textbf{u}\ d^{L_2-1}\textbf{v}\mu(\textbf{u})\mu(\textbf{v})...,
\end{gather}
and the function \(\omega^{s_1,s_2,0}_{L_1,L_2,M}(x_0,\textbf{x}_-,y_0,\textbf{y}_-;z_0)\)
has the following form:

\begin{gather}
\omega^{s_1,s_2,0}_{L_1,L_2,M}(x_0,\textbf{x}_-,y_0,\textbf{y}_-;z_0)=\frac{1}{(2\pi i)^2}\oint\limits_{0}d \eta_x\frac{1}{\eta_x^{s_1+1}}\oint\limits_{0}d \eta_y\frac{1}{\eta_y^{s_2+1}}\omega(x_0,\eta_x\textbf{x}_-,y_0,\eta_y\textbf{y}_-;z_0),\\
\omega(x_0,\textbf{x}_-,y_0,\textbf{y}_-;z_0)=\sum\limits_{\sigma_x,\sigma_y}\frac{1}{|x_{\sigma_x(1)}-z|^2}...\frac{1}{|x_{\sigma_x(L_1-l)}-z|^2}
\frac{1}{|y_{\sigma_y(L_2)}-z|^2}...\frac{1}{|y_{\sigma_y(l+1)}-z|^2}\times\notag\\
\times\frac{1}{|x_{\sigma_x(L_1-l+1)}-y_{\sigma_y(l)}|^2}...\frac{1}{|x_{\sigma_x(L_1)}-y_{\sigma_y(1)}|^2},
\end{gather}
where \(\sigma_x\) and \(\sigma_y\) are cyclic permutations of \(\{1,...,L_1\}\) and \(\{1,...,L_2\}\) correspondingly. Labels \(M\) and \(0\) mean the twist \(M=L_1+L_2-2l\) and spin of operator \(\mathcal{O}_\triangle\) correspondingly.

The three-point correlator of operators with spin is a sum of different tensor structures \cite{Costa:2011mg,Costa:2011dw}. It means that the correlator is characterised by the set of structure constants. As was demonstrated in the \cite{Kazakov:2012ar}(see also Appendix \ref{AppCorWithSpins}),by choosing a special kinematics, one can collapse all those tensor structures into one. To achieve this goal we restrict positions of all local operators to the two-dimension subspace \(\natural=\{n_+,n_-\}\) spanned by two light-ray vectors \(n_+\), \(n_-\). In this case, 3-point correlator has the following form:
\begin{gather}
\langle \mathcal{O}_{L_1,(s_1,\{\alpha_1\})}(x) \mathcal{O}_{L_2,(s_2,\{\alpha_2\})}(y) \mathcal{O}_{\triangle}(z)\rangle_\natural=B_{\natural\left(L_1,(s_1,\{\alpha_1\})\right),\left(L_2,(s_2,\{\alpha_2\})\right),\triangle} \chi(x,y,z),\notag\\
{ } \notag \\
\chi(x,y,z)=\frac{1}
{(x-y)_+^{l}(x-y)_-^{s_1+s_2+l}(x-z)_+^{L_1-l}(x-z)_-^{L_1+s_1-s_2-l}(y-z)_+^{L_2-l}(y-z)_-^{L_2+s_2-s_1-l}},
\end{gather}
where symbol "\(\natural\)" means that we restrict positions of all operators to two-dimension space \(\natural\) .For two-point correlators we get:
\begin{gather}
\langle\mathcal{O}_{L_1,(s_1,\{\alpha_1\})}(x) \bar{\mathcal{O}}_{L_1,(s_1,\{\alpha_1\})}(y)\rangle_\natural=\frac{B_{\natural(L_1,(s_1,\{\alpha_1\}))}}{(x-y)_+^{L_1}(x-y)_-^{L_1+2s_1}},\\
\langle\mathcal{O}_{L_2,(s_2,\{\alpha_2\})}(x) \bar{\mathcal{O}}_{L_2,(s_2,\{\alpha_2\})}(y)\rangle_\natural=\frac{B_{\natural(L_2,(s_2,\{\alpha_2\}))}}{(x-y)_+^{L_2}(x-y)_-^{L_2+2s_2}},\\
\langle\mathcal{O}_{\triangle}(x) \bar{\mathcal{O}}_{\triangle}(y)\rangle_\natural=\frac{B_{\natural \triangle}}{(x-y)_+^{L_1+L_2-2l}(x-y)_-^{L_1+L_2-2l}}.
\end{gather}
The normalized 3-point structure constant is defined as:
\begin{gather}
C_{\natural\left(L_1,(s_1,\{\alpha_1\})\right),\left(L_2,(s_2,\{\alpha_2\})\right),\triangle}=\frac{B_{\natural\left(L_1,(s_1,\{\alpha_1\})\right),\left(L_2,(s_2,\{\alpha_2\})\right),\triangle}}
{\sqrt{B_{\natural(L_1,(s_1,\{\alpha_1\}))}B_{\natural(L_2,(s_2,\{\alpha_2\}))}B_{\natural \triangle}}}.
\end{gather}
As it was mentioned above, one can introduce a normalization factor \(\frac{1}{\mathcal{N}}\), and cancel the contribution of operator \(\mathcal{O}_{\triangle}\). Namely, we specify this factor as a normalized structure constant \(\mathcal{N}=C_{\natural\left(L_1,(0,\{\emptyset\})\right),\left(L_2,(0,\{\emptyset\})\right),\triangle}\) of three operators \(\text{tr}Z^{L_1}\),\(\text{tr}\bar{Z}^{L_2}\) and \(\mathcal{O}_{\triangle}\). It can be easily calculated:
\begin{gather}
\mathcal{N}=C_{\natural\left(L_1,(0,\{\emptyset\})\right),\left(L_2,(0,\{\emptyset\})\right),\triangle}=\frac{1}{N_c}\frac{c_{\triangle}}{2^\frac{L_1+L_2-2l}{2}}\sqrt{\frac{L_1L_2}{N_{\triangle}}},
\end{gather}
where \(N_{\triangle}\) is defined through the 2-point correlator \(\langle\mathcal{O}_{\triangle}(x)\bar{\mathcal{O}}_{\triangle}(y)\rangle_\natural=\frac{N_{\triangle}}{((x-y)_+(x-y)_-)^{L_1+L_2-2l}}\)
The ratio of two normalized 3-point correlators can be expressed in the following way:

\begin{gather}
\frac{C_{\natural\left(L_1,(s_1,\{\alpha_1\})\right),\left(L_2,(s_2,\{\alpha_2\})\right),\triangle}}
{C_{\natural\left(L_1,(0,\{\emptyset\})\right),\left(L_2,(0,\{\emptyset\})\right),\triangle}}  =\frac{2^{\frac{L_1+L_2-2l}{2}}}{\sqrt{L_1L_2}}\frac{\Theta_{(L_1,(s_1,\{\alpha_1\})),(L_2(s_2,\{\alpha_2\}))}(\textbf{x}_-,\textbf{y}_-)}
{\theta_{(L_1,(s_1,\{\alpha_1\}))}(\textbf{x}_-) \theta_{(L_2,(s_2,\{\alpha_2\}))}(\textbf{y}_-)},\label{3point}
\end{gather}
where
\begin{gather}
\Theta_{(L_1,(s_1,\{\alpha_1\})),(L_2,(s_2,\{\alpha_2\}))}(\textbf{x}_-,\textbf{y}_-)=\notag\\
=\langle (L_1,(s_1,\{\alpha_1\}))_{\textbf{u}},(L_2,(s_2,\{\alpha_2\}))_{\textbf{v}}|\ |\mathbb{Q}_{\textbf{x}_-}(\textbf{u})\mathbb{Q}_{\textbf{y}_-}(\textbf{v})\omega^{s_1,s_2,0}_{L_1,L_2,M}(x_0,\textbf{x}_-,y_0,\textbf{y}_-;z_0)\rangle_{\natural\mu}\times\notag\\
(\chi(x_0,y_0,z_0))^{-1},\label{Theta3}\\
\theta_{(L_1,(s_1,\{\alpha_1\}))}(\textbf{x}_-)=\notag\\
=\langle (L_1,(s_1,\{\alpha_1\}))_{\textbf{u}},(L_1,(s_1,\{\alpha_1\}))_{\textbf{v}}|\ |\mathbb{Q}_{\textbf{x}_-}(\textbf{u})\mathbb{Q}_{\textbf{y}_-}(\textbf{v})\omega^{s_1}_{L_1}(x_0,\textbf{x}_-,y_0,\textbf{y}_-)\rangle_{\natural\mu}|_{ \textbf{y}_-\rightarrow \textbf{x}_-}\times\notag\\
\times(x_0-y_0)_+^{L_1}(x_0-y_0)_-^{L_1+2s_1},\\
\theta_{(L_2,(s_2,\{\alpha_1\}))}(\textbf{y}_-)=\notag\\
=\langle (L_2,(s_2,\{\alpha_2\}))_{\textbf{u}},(L_2,(s_2,\{\alpha_2\}))_{\textbf{v}}|\ |\mathbb{Q}_{\textbf{x}_-}(\textbf{u})\mathbb{Q}_{\textbf{y}_-}(\textbf{v})\omega^{s_2}_{L_2}(x_0,\textbf{x}_-,y_0,\textbf{y}_-)\rangle_{\natural\mu}|_{ \textbf{x}_-\rightarrow \textbf{y}_-}\times\notag\\
\times(x_0-y_0)_+^{L_2}(x_0-y_0)_-^{L_2+2s_2}.\label{theta2}
\end{gather}
The last multiplier of all expressions (\ref{Theta3})-(\ref{theta2}) was introduced to cancel coordinate dependence.

\subsection{Case of three $sl(2)$ operator}

In order to construct three-point correlator of three \(sl(2)\) operators, let us introduce three 6-dimension vectors \(p_1\), \(p_2\), \(p_3\) with zero norm \(|p_i|^2=0\) and nonzero pairwise scalar products \(( p_m,p_k )\neq 0\). Then we can introduce three \(sl(2)\) sectors\footnote{We can also choose three different polarizations \(n_m\) in the coordinate space, but for the sake of brevity in notations we restrict ourselves to the case of only one polarization \(n_+\).} which consists of operators with the form \(\mathcal{O}_{L_m,(s_m,\alpha_m)}=\text{tr} D_+^{s_m}U_m^{L_m}\), where \(U_m=\sum\limits_{J=1}^{6}p_m^J\phi^J\) and \(\{\phi^J\}\) are 6 scalar fields in \(N\)=4 SYM. The propagator between two fields \(U_k(x)\) and \(U_m(y)\) has the following form:
\begin{gather}
\langle U_k(x)^a_b U_m(y)^c_d \rangle=\frac{\delta^a_d\delta^c_b}{N_c}\frac{( p_k,p_m )}{|x-y|^2}.
\end{gather}
Thus, we construct nonzero three-point correlator:
\begin{gather}
\Upsilon_{1,2,3}(x,y,z)=\langle\mathcal{O}_{L_1,(s_1,\alpha_1)}(x)\mathcal{O}_{L_2,(s_2,\alpha_2)}(y)\mathcal{O}_{L_3,(s_3,\alpha_3)}(z)\rangle.
\end{gather}
We can obtain the representation similar to (\ref{3-point})
\begin{gather}
\Phi_{L_1,(s_1,\{\alpha_1\})}(\textbf{x}_-) \Phi_{L_2,(s_2,\{\alpha_2\})}(\textbf{y}_-)\Phi_{L_3,(s_3,\{\alpha_3\})}(\textbf{z}_-) \Upsilon_{1,2,3}(x_0,y_0,z_0)=\notag\\
=H_{123}\langle 1_{\textbf{u}},2_{\textbf{v}},3_{\textbf{w}} |\mathbb{Q}_{\textbf{x}_-}(\textbf{u})\mathbb{Q}_{\textbf{y}_-}(\textbf{v})\mathbb{Q}_{\textbf{z}_-}(\textbf{w})\omega^{s_1,s_2,s_3}_{L_1,L_2,L_3}(x_0,\textbf{x}_-;y_0,\textbf{y}_-;z_0,\textbf{z}_-)\rangle_\mu, \label{3sl2}
\end{gather}
where \newline \(H_{123}=\frac{1}{N_c}\frac{c_{L_1,(s_1,\{\alpha_1\})}^{L_1-1}c_{L_2,(s_2,\{\alpha_2\})}^{L_2-1}c_{L_3,(s_3,\{\alpha_3\})}^{L_3-1}}{N_{L_1,(s_1,\{\alpha_1\})}N_{L_2,(s_2,\{\alpha_2\})}N_{L_3,(s_3,\{\alpha_3\})}}
(p_1,p_3)^{\frac{L_3+L_1-L_2}{2}}(p_3,p_2)^{\frac{L_3+L_2-L_1}{2}}(p_2,p_1)^{\frac{L_2+L_1-L_3}{2}}\),
\begin{gather}
\langle 1_{\textbf{u}},2_{\textbf{v}},3_{\textbf{w}}|=\Omega_{L_1,(s_1,\{\alpha_1\})}(\textbf{u})\Omega_{L_2,(s_2,\{\alpha_2\})}(\textbf{v})\Omega_{L_3,(s_3,\{\alpha_3\})}(\textbf{w}),\notag\\
\langle...\rangle_\mu=\int\limits_{R^{L_1-1}}\int\limits_{R^{L_2-1}}\int\limits_{R^{L_3-1}}d^{L_1-1}\textbf{u}\ d^{L_2-1}\textbf{v}d^{L_3-1}\textbf{w}\mu(\textbf{u})\mu(\textbf{v})\mu(\textbf{w})...,
\end{gather}
and the function \(\omega^{s_1,s_2,s_3}_{L_1,L_2,L_3}(\textbf{x}_-,\textbf{y}_-,\textbf{z}_-)\)
has the following form:

\begin{gather}
\omega^{s_1,s_2,s_3}_{L_1,L_2,L_3}(x_0,\textbf{x}_-;y_0,\textbf{y}_-;z_0,\textbf{z}_-)=\notag\\
=\frac{1}{(2\pi i)^3}\oint\limits_{0}d \eta_x\frac{1}{\eta_x^{s_1+1}}\oint\limits_{0}d \eta_y\frac{1}{\eta_y^{s_2+1}}\oint\limits_{0}d \eta_z\frac{1}{\eta_z^{s_3+1}}\omega_{L_1,L_2,L_3}(x_0,\eta_x\textbf{x}_-,y_0;\eta_y\textbf{y}_-;z_0,\eta_z\textbf{z}_-),\\
\omega_{L_1,L_2,L_3}(x_0,\textbf{x}_-;y_0,\textbf{y}_-;z_0,\textbf{z}_-)=\sum\limits_{\sigma_x,\sigma_y,\sigma_z}\frac{1}{|x_{\sigma_x}(1)-z_{\sigma_z}(L_3)|^2...|x_{\sigma_x}(M_2)-z_{\sigma_z}(M_1+1)|^2}\times\notag\\
\times \frac{1}{|x_{\sigma_x}(L_1)-y_{\sigma_y}(1)|^2...|x_{\sigma_x}(M_2+1)-y_{\sigma_y}(M_3)|^2}\frac{1}{|y_{\sigma_y}(L_2)-z_{\sigma_z}(1)|^2...|y_{\sigma_y}(M_3+1)-z_{\sigma_z}(M_1)|^2},
\end{gather}
where \(M_1=\frac{L_2+L_3-L_1}{2}\), \(M_2=\frac{L_1+L_3-L_2}{2}\), \(M_3=\frac{L_1+L_2-L_3}{2}\). \(\sigma_x\), \(\sigma_y\) and \(\sigma_z\)  are cyclic permutations of \(\{1,...,L_1\}\), \(\{1,...,L_2\}\) and \(\{1,...,L_3\}\) correspondingly.
For discussion on how this method works for numerical calculations see Appendix \ref{ForMathematica}.

\subsection{Case of twist-2 operators. Comparing with direct calculation}
In the special case of twist-2 operators one can calculate left-hand side of (\ref{3point}), using explicit form of operators \(\mathcal{O}_{2,(s_1,\{\alpha_1\})}(x)\),\(\mathcal{O}_{2,(s_2,\{\alpha_2\})}(y)\) through the Gegenbauer polinomials:

\begin{gather}
\mathcal{O}_{2,(s_1,\{\alpha_1\})}(x)=\text{tr}(\overrightarrow{D_+}+\overleftarrow{D_+})^{s_1}Z(x)C^{\frac{1}{2}}_{s_1}\left(\frac{\overrightarrow{D_+}-\overleftarrow{D_+}}{\overrightarrow{D_+}+\overleftarrow{D_+}}\right)Z(x),\\
\mathcal{O}_{2,(s_2,\{\alpha_2\})}(y)=\text{tr}(\overrightarrow{D_+}+\overleftarrow{D_+})^{s_2}\bar{Z}(y)C^{\frac{1}{2}}_{s_2}\left(\frac{\overrightarrow{D_+}-\overleftarrow{D_+}}{\overrightarrow{D_+}+\overleftarrow{D_+}}\right)\bar{Z}(y).
\end{gather}

The direct calculation gives us:
\begin{gather}
\frac{C_{\natural\left(2,(s_1,\{\alpha_1\})\right),\left(2,(s_2,\{\alpha_2\})\right),\triangle}}
{C_{\natural\left(2,(0,\{\emptyset\})\right),\left(2,(0,\{\emptyset\})\right),\triangle}} =\frac{(s_1+s_2)!}{\sqrt{(2s_1)!(2s_2)!}}.\label{3pTwist2}
\end{gather}
On the other hand the measure in this case is \(\mu(u)=1\), and \(Q\)-function is Hahn polynomial \(Q_s(u)\sim{}_3F_2(-s,s+1,\frac{1}{2}-iu;1,1;1)\). Using these explicit formulas and Appendix \ref{ForMathematica}, we have checked that the right hand side of (\ref{3point}) exactly coincides with (\ref{3pTwist2}).
\section{Conclusions}

In this paper we have proposed a new approach to the leading order calculation of two- (\ref{2-pointFinal}) and three-point (\ref{3-point}), (\ref{3point}),(\ref{3sl2}) correlation functions of \(sl(2)\) operators. It is important to stress that our construction gives us in one calculation both 2-,3- point correlator and polynomials \(\Phi_{s,\{\alpha\}}\) which are dual to wave functions \(\Psi_{s,\{\alpha\}}\). As the initial data we use only Baxter Q-function. This approach is based on the decomposition (\ref{Decomposition over local}) and Sklyanin's method of separated variables. SoV representation is one of the most general methods in Integrability. We suppose that this approach can be efficiently applied to the study of large spin \(s\) case \cite{Sobko}, because number of integrals doesn't depend on \(s\). It would be interesting to generalize our construction to \(su(2)\) case, and clarify a connection with the method proposed in \cite{Escobedo:2010xs}.

The wave function in SoV representation is constructed from Q-functions. On the other hand, recently proposed \(P-\mu\) system \cite{Gromov:2013pga} gives us information on a variety of Q-functions at any coupling. It would be very tempting to construct SoV representation for \(N\)=4 SYM \cite{GromovIGST}, and then, use it for generalization of our method. Another important point to stress in this context concerns nonlocal light-ray operators (\ref{DefNonLoc}). They are our starting objects, and they preserve their form at any coupling constant.

Our construction can also be useful for the calculation of correlators of generalized operators, such as \(\mathcal{O}_\omega=\text{tr} ZD_+^{-1+\omega}Z\) with \(\omega\rightarrow0\). Operators \(\mathcal{O}_\omega\) play an important role in the BFKL physics, and they were recently understood \cite{Balitsky:2013npa} as nonlocal light-ray operators realizing principal series representation of \(sl(2,R)\). Corresponding noncompact spin-chain can not be solved by ABA technic due
to the absence of the extremal-weight vector. However, it was solved by authors of \cite{Kirch:2004mk}, who have applied both methods of Baxter Q-operator and Separation of Variables.

\section*{Acknowledgments}
 I would like to thank Joao Caetano, Nikolay Gromov, Vladimir Kazakov, Ivan Kostov, Didina Serban, Pedro Vieira and Konstantin Zarembo for interesting discussions on three-point correlation functions in \(N\)=4 SYM. I especially thank Vladimir Kazakov,Gregory Korchemsky and Pedro Vieira for valuable comments concerning this manuscript.  I am especially grateful to Gregory Korchemsky for generously sharing with me his knowledge on the subject in question. I thank Perimeter Institute in Waterloo, Canada,  Kavli IPMU in Tokyo, as well as Yukawa Institute for Theoretical Physics in Kyoto for their hospitality. I thank ERC grant and "Project Unification of Fundamental Forces and Applications" (UNIFY, Number 269217) for support.
\section*{Appendices}

\appendix

\section{Three-point correlator of operators with spins.}\label{AppCorWithSpins}

This appendix is a reminder of the formulas obtained in  the paper \cite{Costa:2011dw}, with some precisions for our particular cases.  According to its methods, a formula for correlation function of any three primary operators with dimensions \(\Delta_i\) and spins \(l_i\) was obtained, using the embedding formalism. Below we give their expression in original notations and apply it to the particular case, when all operators are restricted to two-dimensional plane \(\natural=\{n_+,n_-\}\). Embedding formalism implies the embedding of physical space \(\mathcal{V}=\mathcal{R}^d\) (\(\mathcal{R}^{d-k,k}\)) into the space
\(\mathcal{M}=\mathcal{R}^{1,d+1}\) (\(\mathcal{R}^{d-k+1,k+1}\)) where the conformal group \(SO(1,d+1)\) (\(SO(d-k+1,k+1)\)) is realized linearly. The vector \(x\) from \(\mathcal{V}\) lifts up to \(\mathcal{M}\) by the formula \(x \leftrightarrow P_x=(1,x^2,x)\) , which sets the one-to-one correspondence of vectors from \(\mathcal{V}\) and light-rays in \(\mathcal{M}\). Scalar product of two vectors \(P_1=(P_{1+},P_{1-},p_1)\) and \(P_2\) from \(\mathcal{M}\) sets as \((P_1\cdot P_2)=-\frac{P_{1+}P_{2-}+P_{1-}P_{2+}}{2}+p_1 p_2\), where   \(p_1 p_2\) means the scalar product in \(\mathcal{V}\). In the paper \cite{Costa:2011dw},   three vectors of polarization \(Z_i \leftrightarrow z_i\) were introduced which contract  tensor indices of each operator: \(\phi(x,z)=\phi_{a_1,...,a_l}z^{a_1}...z^{a_l}\). In our case this corresponds to the projection of all indexes on \(n_+\) direction. Thus in our case all indices have the same polarization \(z_1=z_2=z_3=n_+\). The formula for three-point correlation function reads in these notations as follows:

\begin{gather}
\langle\Phi(P_1,Z_{n_+})\Phi(P_2,Z_{n_+})\Phi(P_3,Z_{n_+})\rangle=\sum\limits_{n_{12},n_{13},n_{23}\geq 0}\lambda_{n_{12},n_{13},n_{23}}
\begin{bmatrix}
\Delta_1 & \Delta_2 & \Delta_3\\
l_1 & l_2 & l_3 \\
n_{23}& n_{13}&n_{12}
\end{bmatrix},\label{SumTens}
\end{gather}
where summation goes over all possible tensor structures. The coefficients \(\lambda_{n_{12},n_{13},n_{23}}\) are labeled by the set \(\{n_{12},n_{13},n_{23}\}\) of integers satisfying the following inequalities \(m_1=l_{1}-n_{12}-n_{13}\geq 0 \), \(m_2=l_{2}-n_{12}-n_{23}\geq 0 \), \(m_3=l_{3}-n_{13}-n_{23}\geq 0 \) and the tensor structures are explicitly given by

\begin{gather}
\begin{bmatrix}
\Delta_1 & \Delta_2 & \Delta_3\\
l_1 & l_2 & l_3 \\
n_{23}& n_{13}&n_{12}
\end{bmatrix}=
\frac{V_1^{m_1}V_2^{m_2}V_3^{m_3}H_{12}^{n_{12}}H_{13}^{n_{13}}H_{23}^{n_{23}}}{P_{12}^{\frac{1}{2}(\tau_1+\tau_2-\tau_3)}P_{13}^{\frac{1}{2}(\tau_1+\tau_3-\tau_2)}
P_{23}^{\frac{1}{2}(\tau_2+\tau_3-\tau_1)}},
\end{gather}
where
\begin{gather}
\tau_i=\Delta_i+l_i,\\
P_{ij}=-2(P_i\cdot P_j)=x_{ij}^2,\\
H_{ij}=-2\left((Z_i\cdot Z_j)(P_i\cdot P_{j})-(Z_i\cdot P_j)(Z_j\cdot P_{i})\right)=-2x_{ij+}^2,\\
V_{i,jk}=\frac{(Z_i\cdot P_j)(P_i\cdot P_k)-(Z_i\cdot P_k)(P_i\cdot P_j)}{(P_j\cdot P_k)},\\
V_1=V_{1,23}=\frac{x_{21+}x_{13}^2-x_{31+}x_{12}^2}{x_{23}^2}, \\
V_2=V_{2,31}=\frac{x_{32+}x_{21}^2-x_{12+}x_{23}^2}{x_{13}^2}, \\
V_3=V_{3,12}=\frac{x_{13+}x_{23}^2-x_{23+}x_{13}^2}{x_{12}^2}.
\end{gather}

In a general case all tensor structures are different. In the case  when the coordinates are restricted to the \(\{n_+,n_-\}\) - plane we get much simpler expressions for \(V_i\):

\begin{gather}
V_1=-\frac{x_{12+}x_{13+}}{x_{23+}},\ V_2=-\frac{x_{23+}x_{12+}}{x_{13+}},\ V_3=-\frac{x_{13+}x_{23+}}{x_{12+}}.
\end{gather}
In this case all tensor structures are collapsed in one:

\begin{gather}
\begin{bmatrix}
\Delta_1 & \Delta_2 & \Delta_3\\
l_1 & l_2 & l_3 \\
n_{23}& n_{13}&n_{12}
\end{bmatrix}=(-1)^{l_1+l_2+l_3-n_{12}-n_{13}-n_{23}}2^{n_{12}+n_{13}+n_{23}-\frac{1}{2}(\tau_1+\tau_2+\tau_3)}\times\notag\\
\times \frac{1}{x_{12-}^{a(1,2|3)}x_{12+}^{b(1,2|3)}x_{13-}^{a(1,3|2)}x_{13+}^{b(1,3|2)}x_{23-}^{a(2,3|1)}x_{23+}^{b(2,3|1)}},
\end{gather}
where

\begin{gather}
a(i,j|k)=\frac{1}{2}(\Delta_i+l_i+\Delta_j+l_j-\Delta_k-l_k),\\
b(i,j|k)=\frac{1}{2}(\Delta_i-l_i+\Delta_j-l_j-\Delta_k+l_k).
\end{gather}

\section{The method at work}\label{ForMathematica}
Our method can be easily realized in any mathematical package, such as \textit{Mathematica}. Indeed, let us look through the main steps of calculation in (\ref{2-pointFinal}), (\ref{3-point}), (\ref{3point}) or (\ref{3sl2}).

\textbf{Step 1}. As a first step, we should calculate functions  \(\omega^{s}_{L}(x_0,\textbf{x}_-,y_0,\textbf{y}_-)\),
\newline
\(\omega^{s_1,s_2,0}_{L_1,L_2,M}(x_0,\textbf{x}_-,y_0,\textbf{y}_-;z_0)\), \(\omega^{s_1,s_2,s_3}_{L_1,L_2,L_3}(x_0,\textbf{x}_-;y_0,\textbf{y}_-;z_0,\textbf{z}_-)\). This calculation corresponds to the calculation of residue of an rational function. It can be easily carried out by \textit{Mathematica}.

\textbf{Step 2}.  The next step is to act by operators \(\mathbb{Q}_{\textbf{x}_-}(\textbf{u})\) which are defined in (\ref{ActionQ}). After proceeding through \textbf{Step 1} we get polynomial expression w.r.t. \(\textbf{x}_-\). It is easy to see that all integrals which can appear, have very simple form:
\begin{gather}
\int\limits_0^1 d\tau_{ij} \tau_{ij}^{-iu_i-\frac{1}{2}+k}(1-\tau_{ij})^{iu_i-\frac{1}{2}+m}=B(-iu_i+\frac{1}{2}+k,iu_i+\frac{1}{2}+m),
\end{gather}
where \(k\) and \(m\) are integer numbers. One can rewrite Beta-function in the following way:
\begin{gather}
B(-iu_i+\frac{1}{2}+k,iu_i+\frac{1}{2}+m)=\frac{1}{(k+m)!}\frac{\pi}{\cosh(\pi u_i)}\frac{\Gamma(-iu_i+\frac{1}{2}+k)}{\Gamma(-iu_i+\frac{1}{2})}\frac{\Gamma(iu_i+\frac{1}{2}+m)}{\Gamma(iu_i+\frac{1}{2})}.
\end{gather}
Terms \(\frac{\Gamma(-iu_i+\frac{1}{2}+k)}{\Gamma(-iu_i+\frac{1}{2})}\) and \(\frac{\Gamma(iu_i+\frac{1}{2}+m)}{\Gamma(iu_i+\frac{1}{2})}\) are polynomials w.r.t. \(u_i\). It means that  acting by \(\mathbb{Q}_{\textbf{x}_-}(\textbf{u})\) we get the following expression:
\begin{gather}
P(\textbf{u})\prod\limits_{i=1}^{L-1}\frac{1}{(\cosh(\pi u_i))^L},
\end{gather}
where \(P(\textbf{u})\) is a polynomial of variables \((u_1,...,u_{L-1})\).

\textbf{Step 3}. Finally, we should carry out an integration with measure \(\mu\) (\ref{MeasurFin}). It is easy to see, that all integrals have the following form:

\begin{gather}
I(L,k,m)=\int\limits_{-\infty}^{\infty}d u_i \frac{u_i^k e^{m \pi u_i}}{(\cosh{\pi u_i})^{L}}.
\end{gather}
Number \(m\) is less then \(L\) because the power of the exponent comes from the measure (\ref{MeasurFin}). It means that, all integrals are well defined. This integral can be calculated by \textit{Mathematica} for any integer numbers \(k,m,L\). Moreover, one can reduce this integral to the calculation of derivatives:

\begin{gather}
I(L,k,m)=\frac{1}{\pi^k}\frac{\partial^k}{\partial m^k}I(L,0,m),\\
I(L,0,m)=\frac{2^L}{L+m} { }_2F_1(L,\frac{L+m}{2},1+\frac{L+m}{2},-1)+(m\leftrightarrow-m)
\end{gather}

\printindex

\end{document}